\documentclass[
showpacs,preprintnumbers,amsmath,amssymb]{revtex4}
\usepackage{amsmath}

\begin{document}

\tolerance=5000

\def\pp{{\, \mid \hskip -1.5mm =}}
\def\cL{{\cal L}}
\def\be{\begin{equation}}
\def\ee{\end{equation}}
\def\bea{\begin{eqnarray}}
\def\eea{\end{eqnarray}}
\def\beq{\begin{eqnarray}}
\def\eeq{\end{eqnarray}}
\def\tr{{\rm tr}\, }
\def\nn{\nonumber \\}
\def\e{{\rm e}}

\title{Unifying inflation with LambdaCDM epoch in modified f(R) gravity
consistent with Solar System tests}
\author{Shin'ichi Nojiri}
\email{nojiri@phys.nagoya-u.ac.jp}
\affiliation{Department of Physics, Nagoya University, Nagoya 464-8602. Japan}
\author{Sergei D. Odintsov\footnote{also at Lab. Fundam. Study, Tomsk State
Pedagogical University, Tomsk}}
\email{odintsov@ieec.uab.es}
\affiliation{Instituci\`{o} Catalana de Recerca i Estudis Avan\c{c}ats (ICREA)
and Institut de Ciencies de l'Espai (IEEC-CSIC),
Campus UAB, Facultat de Ciencies, Torre C5-Par-2a pl, E-08193 Bellaterra
(Barcelona), Spain}

\begin{abstract}

We suggest two realistic $f(R)$ and one $F(G)$ modified gravities which
are consistent with local tests and cosmological bounds. The typical
property of such theories is the presence of the effective cosmological
constant epochs in such a way that early-time inflation and late-time
cosmic acceleration are naturally unified within single model. It is shown
that classical instability does not appear here and Newton law is
respected. Some discussion of possible anti-gravity regime appearence and
related modification of the theory is
done.

\end{abstract}


\maketitle

\noindent
1. It is expected that modified gravity (for a review, see \cite{review})
approach which may be related with string/M-theory\cite{string} may help
in the understanding of the dark energy origin. For instance, various
modified $f(R)$ gravities have been constructed and applied to the
description of the late-time cosmic acceleration
\cite{CDTT,NO,FR,FR1,cap,lea} with also the check of local tests.
Even the
form of modified $f(R)$ gravity may be reconstructed from the known
universe expansion history\cite{recrev}. Hence, such gravitational
alternative for dark energy may be considered as an alternative gravity
theory subject that it passes cosmological bounds and Solar System tests.

Recently, it has been proposed the class of modified gravities with the
effective cosmological constant epoch \cite{cap,HS,no,AB,star}. The very
simple versions of such theories\cite{HS,no,AB,star} with vanishing
cosmological constant may describe successfully the $\Lambda$CDM epoch
and pass the local tests/cosmological bounds. Nevertheless,
the early universe epoch is not included there and deviations from Newton
law at large scales may occur.

In the present letter we suggest two realistic new models of modified
$f(R)$
gravity which may describe the early-time inflation and late-time
acceleration in the unified manner, extending the earlier proposal of
ref.\cite{NO}. Moreover, such theories successfully pass the Solar System
tests as well as simplest cosmological bounds and they are free of
instabilities. Nevertheless, the first of suggested models may develop
the anti-gravity regime, that is why  its simple
modification to
avoid this problem is done. Finally, viable $F(G)$ gravity which also
unifies
inflation with late-time acceleration and may pass the
local tests/cosmological bounds is suggested.

\

\noindent
2. The action of general $f(R)$ gravity is given by
\be
\label{XXX7}
S=\frac{1}{\kappa^2}\int d^4 x \sqrt{-g} \left(R + f(R)\right)\ .
\ee
Here $f(R)$ is an arbitrary function.
The  equation of motion  in $f(R)$-gravity with matter is given by
\be
\label{XXX22}
\frac{1}{2}g_{\mu\nu} F(R) - R_{\mu\nu} F'(R) - g_{\mu\nu} \Box F'(R)  + \nabla_\mu \nabla_\nu F'(R)
= - \frac{\kappa^2}{2}T_{(m)\mu\nu}\ .
\ee
Here $F(R)=R+f(R)$ and $T_{(m)\mu\nu}$ is the matter energy-momentum tensor.

Recently an interesting $f(R)$ model has been proposed in \cite{HS}. In the model $f(R)$ is given by
\be
\label{HS1}
f_{HS}(R)=-\frac{m^2 c_1 \left(R/m^2\right)^n}{c_2 \left(R/m^2\right)^n + 1}\ ,
\ee
which satisfies the condition
\bea
\label{HS2}
\lim_{R\to\infty} f_{HS} (R) &=& \mbox{const}\ ,\nn
\lim_{R\to 0} f_{HS}(R) &=& 0\ ,
\eea
The second condition means that there could be a flat spacetime solution
(vanishing cosmological constant).
The estimation of ref.\cite{HS} suggests that $R/m^2$ is not so small but rather large even
in the present universe and $R/m^2\sim 41$.
Hence,
\be
\label{HSb1}
f_{HS}(R)\sim - \frac{m^2 c_1}{c_2} + \frac{m^2 c_1}{c_2^2} \left(\frac{R}{m^2}\right)^{-n}\ ,
\ee
which gives an ``effective'' cosmological constant $-m^2 c_1/c_2$ and
generates the late-time accelerating
expansion.
One can show that
\be
\label{HSbb1}
H^2 \sim \frac{m^2 c_1 \kappa^2 }{c_2} \sim \left(70 \rm{km/s\cdot pc}\right)^2 \sim \left(10^{-33}{\rm eV}\right)^2\ .
\ee
At the intermediate epoch, where the matter density $\rho$ is larger
than the effective cosmological constant,
\be
\label{HSbb2}
\rho > \frac{m^2 c_1}{c_2}\ ,
\ee
there appears the matter dominated phase and the universe
expands with deceleration. Hence, above model describes the effective
$\Lambda$CDM cosmology.

Although the model \cite{HS} is very succesful (see, however,
ref.\cite{no} for
description of the Newton law at large scales here), the early
time inflation  is not included there. We suggest the modified gravity
model to treat the
inflation
 and the late-time accelerating expansion in a unified way.
We now consider simple extention of the model \cite{HS}
 to include the inflation at the early universe.
In order to generate the inflation, one may require
\be
\label{Uf1}
\lim_{R\to\infty} f (R) = - \Lambda_i\ .
\ee
Here $\Lambda_i$ is an effective cosmological constant at the early
universe and therefore
we assume $\Lambda_i \gg \left(10^{-33}{\rm eV}\right)^2$.
We may assume $\Lambda_i\sim 10^{20\sim38}$.
Instead of (\ref{Uf1}), as in Starobinsky's model, one may require
\be
\label{Uf2}
\lim_{R\to\infty} f (R) \propto R^2 \ .
\ee
In order that the accelerating expansion in the present universe could be
 generated, let us consider that
$f(R)$ could be a small constant at  present universe, that is,
\be
\label{Uf3}
f(R_0)= - 2R_0\ ,\quad
f'(R_0)\sim 0\ .
\ee
Here $R_0$ is current curvature  $R_0\sim \left(10^{-33}{\rm
eV}\right)^2$.
The next condition corresponding to the second one in (\ref{HS2}) is:
\be
\label{Uf4}
\lim_{R\to 0} f(R) = 0\ .
\ee
In the above class of models, the universe  starts from the inflation
driven by the effective cosmological
constant (\ref{Uf1}) at the early stage, where curvature is very
large. As curvature becomes smaller, the
effective cosmological constant also becomes smaller. After that the
radiation/matter  dominates.
When the density of the radiation and the matter becomes small and the
curvature goes to the value $R_0$
(\ref{Uf3}), there  appears the small effective cosmological constant
(\ref{Uf3}). Hence, the current cosmic
expansion could start.

An example satisfying (\ref{Uf1}), (\ref{Uf3}), and (\ref{Uf4}) is
\be
\label{Uf5}
f(R) = - \frac{\left(R-R_0\right)^{2n+1} + R_0^{2n+1}}{f_0
+ f_1 \left\{\left(R-R_0\right)^{2n+1} + R_0^{2n+1}\right\}}
=- \frac{1}{f_1} + \frac{f_0/f_1}{f_0
+ f_1 \left\{\left(R-R_0\right)^{2n+1} + R_0^{2n+1}\right\}}\ .
\ee
Here $n$ is a positive integer, $n=1,2,3,\cdots$ and
\be
\label{Uf6}
\frac{R_0^{2n+1}}{f_0 +f_1  R_0^{2n+1}}=2R_0\ ,\quad
\frac{1}{f_1} = \Lambda_i\ ,
\ee
that is
\be
\label{Uf7}
f_0=\frac{R_0^{2n}}{2} - \frac{R_0^{2n+1}}{\Lambda_i}
\sim \frac{R_0^{2n}}{2} \ ,\quad
f_1=\frac{1}{\Lambda_i}\ .
\ee

By introducing the auxilliary field $A$ one may rewrite the action
(\ref{XXX7}) in the following form:
\be
\label{XXX10}
S=\frac{1}{\kappa^2}\int d^4 x \sqrt{-g} \left\{\left(1+f'(A)\right)\left(R-A\right) + A + f(A)\right\}\ .
\ee
 From the equation of motion with respect to $A$, it follows $A=R$.
By using the scale transformation $g_{\mu\nu}\to \e^\sigma g_{\mu\nu}$ with $\sigma = -\ln\left( 1 + f'(A)\right)$,
we obtain the Einstein frame action:
\bea
\label{XXX11}
S_E &=& \frac{1}{\kappa^2}\int d^4 x \sqrt{-g} \left\{ R - \frac{3}{2}\left(\frac{F''(A)}{F'(A)}\right)^2
g^{\rho\sigma}\partial_\rho A \partial_\sigma A - \frac{A}{F'(A)}
+ \frac{F(A)}{F'(A)^2}\right\} \nn
&&=\frac{1}{\kappa^2}\int d^4 x \sqrt{-g} \left( R - \frac{3}{2}g^{\rho\sigma}
\partial_\rho \sigma \partial_\sigma \sigma - V(\sigma)\right)\ , \\
V(\sigma) &=& \e^\sigma g\left(\e^{-\sigma}\right) - \e^{2\sigma} f\left(g\left(\e^{-\sigma}\right)\right)
= \frac{A}{F'(A)} - \frac{F(A)}{F'(A)^2}\ .
\eea
Here $g\left(\e^{-\sigma}\right)$ is given by solving $\sigma = -\ln\left( 1 + f'(A)\right)=\ln F'(A)$
as $A=g\left(\e^{-\sigma}\right)$.
After the scale transformation  $g_{\mu\nu}\to \e^\sigma g_{\mu\nu}$,
there appears a coupling of the scalar field $\sigma$
with the matter. For example, if the matter is the scalar field $\Phi$ with mass $M$, whose action is given by
\be
\label{MN1}
S_\phi=\frac{1}{2}\int d^4x\sqrt{-g}\left(-g^{\mu\nu}\partial_\mu\Phi \partial_\nu\Phi - M^2 \Phi^2\right)\ ,
\ee
there appears a coupling with $\sigma$ in the Einstein frame:
\be
\label{MN2}
S_{\phi\, E}=\frac{1}{2}\int d^4x\sqrt{-g} \left(-\e^{\sigma} g^{\mu\nu}\partial_\mu\Phi \partial_\nu\Phi
 - M^2 \e^{2\sigma}\Phi^2\right)\ .
\ee
The strength of the coupling is  of the gravitational coupling $\kappa$
order. Unless the mass of $\sigma$,
which is defined by
\be
\label{MN3}
m_\sigma^2 \equiv \frac{1}{2}\frac{d^2 V(\sigma)}{d\sigma^2}
=\frac{1}{2}\left\{\frac{A}{F'(A)} - \frac{4F(A)}{\left(F'(A)\right)^2} + \frac{1}{F''(A)}\right\}
\ee
is large, there appears the large correction to the Newton law.

In air on the earth, the scalar curvature could be given by $A=R\sim 10^{-50}\,{\rm eV}^2$.
On the other hand, in the solar system, we find $A=R\sim 10^{-61}\,{\rm eV}^2$.
In the model (\ref{Uf5}) with (\ref{Uf6}), if $n\geq 3$ in the air or $n\geq 10\sim 12$ (if
$\Lambda_i\sim 10^{20\sim 38}$) in the solar system,
we find
\be
\label{Uf8}
f_0 \ll f_1 \left\{\left(R-R_0\right)^{2n+1} + R_0^{2n+1}\right\}
\sim f_1 R^{2n+1}
\ee
and
\be
\label{Uf9}
F(R)=R+f(R)\sim R - \frac{1}{f_1} + \frac{f_0}{f_1^2 R^{2n+1}} \sim \frac{1}{f_1}\ .
\ee
Furthermore, if $n>6 \sim 7$ (if $\Lambda_i\sim 10^{20\sim 38}$) in the air, we find $F'(R)\sim 1$ and
\be
\label{Uf10}
m_\sigma^2 \sim - \frac{F(A)}{F'(A)^2}\sim \frac{2}{f_1} \sim 10^{38\sim 56}\,{\rm eV}^2\ ,
\ee
which is very large and there is no observable correction to the Newton
law.
We should note $1$ mm $\sim (10^{-4}\,{\rm eV})^{-1}$.
On the other hand, in the Solar System, if $n\gg 10\sim 12$ (if
$\Lambda_i\sim 10^{20\sim 38}$) ,
\be
\label{Uf11}
m_\sigma^2 \sim \frac{1}{2F''(A)}\sim \frac{f_0}{f_1 R^{2n+3}} \sim 10^{239\sim 295 - 10n}\,{\rm eV}^2\ ,
\ee
which is large enough since the radius of earth is about $10^7\,{\rm m}\sim 10^{-14}\,{\rm eV}$) and
there is no  visible correction to the Newton law.

There may exist another type of instability (so-called matter instability)
in $f(R)$ gravity \cite{DK}(see, also \cite{Faraoni,no}).
The instability might occur when the curvature is rather large, as on the planet, compared with
the average curvature at the universe $R\sim \left(10^{-33}\,{\rm
eV}\right)^2$.
By multipling Eq.(\ref{XXX22}) with $g^{\mu\nu}$, one obtains
\be
\label{XXX23}
\Box R + \frac{F^{(3)}(R)}{F^{(2)}(R)}\nabla_\rho R \nabla^\rho R
+ \frac{F'(R) R}{3F^{(2)}(R)} - \frac{2F(R)}{3 F^{(2)}(R)}
= \frac{\kappa^2}{6F^{(2)}(R)}T\ .
\ee
Here $T\equiv T_{(m)\rho}^{\ \rho}$ and $F^{(n)}(R) \equiv d^nF(R)/dR^n$.
We consider a perturbation from the following solution of the Einstein
gravity:
\be
\label{XXX24}
R=R_b\equiv - \frac{\kappa^2}{2}T>0\ .
\ee
Note that $T$ is negative since $|p|\ll \rho$ on the earth and
$T=-\rho + 3 p \sim -\rho$. Then we assume
\be
\label{XXX25}
R=R_b + R_p\ ,\quad \left(\left|R_p\right|\ll \left|R_b\right|\right)\ .
\ee
Now one can get
\bea
\label{XXX26}
0&=&\Box R_b + \frac{F^{(3)}(R_b)}{F^{(2)}(R_b)}\nabla_\rho R_b \nabla^\rho R_b + \frac{F'(R_b) R_b}{3F^{(2)}(R_b)}
 - \frac{2F(R_b)}{3 F^{(2)}(R_b)} - \frac{R_b}{3F^{(2)}(R_b)} \nn
&& + \Box R_p + 2\frac{F^{(3)}(R_b)}{F^{(2)}(R_b)}\nabla_\rho R_b \nabla^\rho R_p + U(R_b) R_p\ , \nn
U(R_b)&\equiv& \left(\frac{F^{(4)}(R_b)}{F^{(2)}(R_b)} - \frac{F^{(3)}(R_b)^2}{F^{(2)}(R_b)^2}\right)
\nabla_\rho R_b \nabla^\rho R_b + \frac{R_b}{3} \nn
&& - \frac{F^{(1)}(R_b) F^{(3)}(R_b) R_b}{3 F^{(2)}(R_b)^2} - \frac{F^{(1)}(R_b)}{3F^{(2)}(R_b)}
+ \frac{2 F(R_b) F^{(3)}(R_b)}{3 F^{(2)}(R_b)^2} - \frac{F^{(3)}(R_b) R_b}{3 F^{(2)}(R_b)^2}
\ .
\eea
Since $\Box R_p \sim - \partial_t^2 R_p$,
Eq.(\ref{XXX26}) has the following form:
\be
\label{XXXXX1}
0=-\partial_t^2 R_p + U(R_b) R_p + {\rm const.}\ .
\ee
Then if $U(R_b)$ is positive, $R_p$ becomes exponentially large as a function of $t$:
$R_p\sim \e^{\sqrt{U(R_b)} t}$
and the system becomes unstable.
In the model (\ref{Uf5}) with (\ref{Uf6}), if $n>2$, we find
\be
\label{Uf12}
U(R_b)\sim - \frac{(2n+3)f_1 R_b^{2n+2}}{3(2n+1)f_0}<0\ .
\ee
Therefore there is no this kind of the instability.

Hence, in the model (\ref{Uf5}), the universe could start from the
inflation. As curvature becomes smaller, the
effective cosmological constant  becomes small and
after that the radiation/matter  could dominate.
When the density of the radiation and, later, of the matter becomes small
and
the curvature goes to the value $R_0$,
there could appear the small effective cosmological constant and
 the late-time accelerating expansion starts.

\

\noindent
3. As is clear from (\ref{XXX10}), if $F'(R)=1+f'(R)>0$, the square of the
effective gravitational coupling
becomes negative $\kappa_{\rm eff}^2 \equiv \kappa^2 / F'(A)$ and
 theory enters the anti-gravity regime. Nevertheless, if the theory enters
such regime at the future, such theory may still be viable.
We now check if the model (\ref{Uf5}) could pass the anti-gravity
constraint.
Since
\bea
\label{Uf13}
f'(R) &=& - \frac{(2n+1) f_0 \left(R-R_0\right)^{2n}}{\left(f_0
+ f_1 \left\{\left(R-R_0\right)^{2n+1} + R_0^{2n+1}\right\}\right)^2}\ ,\nn
f''(R) &=& - \frac{2(2n+1) f_0 \left(R-R_0\right)^{2n-1}
\left( - f_0 - R_0^{2n+1} + f_1 (n+1) \left(R-R_0\right)^{2n+1}\right)}{\left(f_0
+ f_1 \left\{\left(R-R_0\right)^{2n+1} + R_0^{2n+1}\right\}\right)^3}\ ,
\eea
$f'(R)$ has a maximum at $R=\tilde R$, which satisfies
\be
\label{Uf14}
\left(\tilde R - R_0 \right)^{2n+1}=\frac{f_0 + f_1 R_0^{2n+1}}{f_1(n+1)} \sim \frac{f_0}{f_1 (n+1)}\ .
\ee
Then the maximum value of $f'(R)$ is given by
\be
\label{Uf15}
f'(\tilde R) \sim - \frac{f_0^{-\frac{1}{2n+1}}}{f_1^{\frac{2n}{2n+1}}} \sim (R_0f_1)^{-\frac{2n}{2n+1}}\ll -1\ .
\ee
Here we have used (\ref{Uf7}). Therefore $F'(R)<0$ and
the anti-gravity regime could occur in general.
We can consider only the region where $F'(R)>0$ or $f'(R)>-1$ without the
 the anti-gravity problem. From (\ref{Uf14}), however, one gets
\be
\label{Uf15b}
\tilde R \sim \left(10^{-33 + \frac{85\sim 91}{2n+1}}{\rm eV}\right)^2\ ,
\ee
which could be small compared with the curvature at the early universe.
There are two solutions for the equation $f'(R)=-1$.
Let denote the larger one by $R_+$ and smaller one by $R_-$. By assuming
\be
\label{Uf15c}
f_0 \ll f_1 \left\{\left(R_+-R_0\right)^{2n+1} + R_0^{2n+1}\right\} \sim f_1 R^{2n+1}\ ,
\ee
we find
\be
\label{Uf15d}
R_+ \sim \left(\frac{(2n+1)f_0}{f_1^2}\right)^{1/(2n+1)} \sim \left(10^{-33 + \frac{208\sim 232}{2n+1}}{\rm eV}\right)^2\ .
\ee
On the other hand, by assuming
\be
\label{Uf15e}
f_0 \gg f_1 \left\{\left(R_--R_0\right)^{2n+1} + R_0^{2n+1}\right\} \ ,
\ee
we find
\be
\label{Uf15f}
R_- \sim R_0 \sim \left(10^{-33}{\rm eV}\right)^2\ .
\ee
Then $R_\pm$ could be rather small compared with the scale of the inflation $\left(10^{19\sim 25}{\rm eV}\right)^2$.
Then the model (\ref{Uf5}) itself seems to be not  viable.
It could be possible to replace $f'(R)$ given from (\ref{Uf5}) in a region $R_- - \epsilon_- < R < R_+ + \epsilon_+$,
where $\epsilon_\pm$ are small positive constants, with a proper function whose first derivative is always
greater than $-1$. For example,
\bea
\label{Uf15g}
f(R)&=&\left\{
\begin{array}{ll}
f_{\rm old}(R) \quad & R<R_- - \epsilon_- \\
f_{\rm old}(R_- - \epsilon_-) + f_{\rm old}'(R_- - \epsilon_-)\left(R - R_- + \epsilon_-\right) \quad
& R_- - \epsilon_- < R < R_+ \epsilon_+ \\
f_{\rm old}(R) + f_{\rm old}'(R_- - \epsilon_-)\left(R_+ + \epsilon_+ - R_- + \epsilon_-\right)
 - f_{\rm old}(R_+ - \epsilon_+) + f_{\rm old}(R_- - \epsilon_-)
\quad
& R > R_+ \epsilon_+ \\
\end{array}\right. \ .\nn
f_{\rm old} &\equiv& - \frac{\left(R-R_0\right)^{2n+1} + R_0^{2n+1}}{f_0
+ f_1 \left\{\left(R-R_0\right)^{2n+1} + R_0^{2n+1}\right\}} \ .
\eea
We now choose $\epsilon_\pm$ to be $f_{\rm old}'(R_- - \epsilon_-)=f_{\rm old}'(R_+ + \epsilon_+)$.
Hence, one obtains $f'(R)>-1$ and there is no problem of anti-gravity so
such corrected model seems to be quite realistic.

 From another side, in order that the model (\ref{Uf5}) could be viable,
there should occur the discrete transition from $R=R_+$ and $R=R_-$.
Such a transition might occur as in the usual phase transition
 with the jump of the value of the order parameter. The properties of such
possible transition should be studied in detail.

\

\noindent
4. In order to avoid the above anti-gravity problem from the very
beginning, we may propose
another model given by
\be
\label{Uf16}
f(R)=-f_0 \int_0^R dR \e^{-\frac{\alpha R_1^{2n}}{\left(R - R_1\right)^{2n}} - \frac{R}{\beta\Lambda_i}}\ .
\ee
Here $\alpha$, $\beta$, $f_0$, and $R_1$ are constants.
Then by construction, as long as $0<f_0<1$, $f'(R)>-1$ and therefore $F'(R)>0$.
Since
\bea
\label{Uf17}
f(R_1) &\sim& -f_0 \int_0^{R_1} dR \e^{-\frac{\alpha R_1^{2n}}{\left(R - R_1\right)^{2n}}}
= -f_0 A_n(\alpha) R_1\ , \nn
A_n(\alpha) &\equiv& \int_0^1 dx \e^{-\frac{\alpha}{x^{2n}}}\ ,
\eea
and $- f(R_1)$ could be identified with the effective cosmological constant $2R_0$, we find
\be
\label{Uf18}
f_0 A_n(\alpha) R_1 = R_0\ .
\ee
Note that $A_n(0)=1$, $A_n(+\infty)=0$, and $A'(x)<0$.
On the other hand, since
\be
\label{Uf19}
f(+\infty) \sim \int_0^{\infty}dR \e^{-{R}{\beta \Lambda_i}} = - f_0 \beta \Lambda_i\ ,
\ee
and $-f(+\infty)$ could be identified with the effective cosmological constant
 at the inflationary epoch, $\Lambda_i$,
we find
\be
\label{Uf19B}
f_0\beta=1\ .
\ee

Let us now investigate the correction to the Newton law.
One gets
\be
\label{Uf20}
f''(R)= -f_0 \e^{-\frac{\alpha R_1^{2n}}{\left(R - R_1\right)^{2n}} - \frac{R}{\beta\Lambda_i}}
\left(\frac{2nR_1^{2n}}{(R- R_1)^{2n+1}} - \frac{1}{\beta\Lambda_i}\right)\ .
\ee
If in air on the earth $n>3\sim 4$ (if $\Lambda_i\sim 10^{20\sim 38}$) or
 in  Solar System
$n>10\sim 12$, we find $m_\sigma^2$  (\ref{MN3}) is given by
\be
\label{Uf21}
m_\sigma^2 \sim \frac{1}{2F''(R)} \sim \frac{\beta\Lambda_i}{2f_0} \sim \left(10^{10\sim 16}\,{\rm GeV}\right)^2\ ,
\ee
which is very large and positive and therefore the correction
to the Newton law is very small.

We now also investigate the instability indicated in \cite{DK}.
Inside the earth, $U(R)$  (\ref{XXX26}) has the following form
\be
\label{Uf22}
U(R_b)\sim - \frac{F^{(0)}(R_b)}{3F^{(2)}(R_b)} \sim - \frac{\beta\Lambda_i}{f_0}<0\ .
\ee
Here we have assumed $n>10\sim 12$. Since $U(R_b)$ is negative,
there does not occur such the instability.

Hence, in the model (\ref{Uf16}), the universe could start from the
inflation driven by the effective cosmological
constant $-f(+\infty)$  at the early stage, where curvature is
large. As curvature becomes small, the
effective cosmological constant  becomes small too and
radiation/matter  dominates at the intermediate epoch.
When the density of the radiation/matter becomes small and the curvature
goes to the value $R_1$,
there  appears the small effective cosmological constant $-f(R_1)$, so
that the accelerating
expansion  starts. Hence, the realistic
modified gravity consistent with Solar System tests and unifying inflation
with cosmic acceleration is constructed.

Instead of $f(R)$-gravity, we may consider the $F(G)$-gravity, where
the action is given by \cite{nojiri}
\be
\label{Uf23}
S=\frac{1}{\kappa^2}\int d^4 x \sqrt{-g} \left(R + F(G)\right)\ ,
\ee
and $G$ is the Gauss-Bonnet invariant:
\be
\label{Uf24}
G=R^2 - 4R_{\mu\nu} R^{\mu\nu} + R_{\mu\nu\rho\sigma}R^{\mu\nu\rho\sigma}\ .
\ee
Note that in $F(G)$-gravity, there are no  problems\cite{nojiri} with the
Newton
law, instabilities and the anti-gravity regime.
One may consider the model similar to (\ref{Uf5}):
\be
\label{Uf25}
F(G) = - \frac{\left(G-G_0\right)^{2n+1} + G_0^{2n+1}}{F_0
+ F_1 \left\{\left(G-G_0\right)^{2n+1} + G_0^{2n+1}\right\}}
=- \frac{1}{F_1} + \frac{F_0/F_1}{f_0
+ F_1 \left\{\left(G-G_0\right)^{2n+1} + G_0^{2n+1}\right\}}\ .
\ee
Here $G_0$ corresponds to the present value of the Gauss-Bonnet invariant.
Since $F'(G)=0$ when $G=G_0$ and $G=+\infty$, $F(G)$ becomes  almost
constant and can be regarded as
the effective cosmological constant. As in (\ref{Uf6}),
 we may identify $F(\infty)$ as the cosmological constant
for the inflationary epoch and $F(R_0)$ as that at the present
accelerating era. Then instead of (\ref{Uf7}),
one gets
\be
\label{Uf26}
F_0=\frac{G_0^{2n}}{2} - \frac{G_0^{2n+1}}{\Lambda_i}
\sim \frac{G_0^{2n}}{2} \ ,\quad
F_1=\frac{1}{\Lambda_i}\ .
\ee
Hence, the universe  starts from the inflation driven by the effective
cosmological
constant $\Lambda_i$  (\ref{Uf26}) at the early epoch, where curvature and
therefore the Gauss-Bonnet invariant $G$
is very large (for the study of cosmological perturbations in such
theory, see \cite{barrow}). As curvature becomes small, the
effective cosmological constant  becomes small too and radiation/matter
 could dominate at the intermediate universe.
When the density of the radiation/matter becomes small and $G$ goes to the
value $G_0$ in
(\ref{Uf25}), there  appears the small effective cosmological constant
$-F(G_0)$, and cosmic acceleration starts.

Thus, we suggested several realistic models of $f(R)$ and $F(G)$ gravity
which propose natural unification of the early-time inflation and
late-time acceleration being consistent with local tests and cosmological
bounds. Definitely, more precise local tests/cosmological checks for such
theories should be made in future, having in mind that more precise
observational data will be available soon.

\ 

\noindent
{\bf Acknowledgements}
The research by S.N. has been supported in part by the
Ministry of Education, Science, Sports and Culture of Japan under
grant no.18549001
and 21st Century COE Program of Nagoya University
provided by Japan Society for the Promotion of Science (15COEG01).
The research by S.D.O. has been supported in part by the projects
FIS2006-02842, FIS2005-01181 (MEC, Spain), by the project 2005SGR00790
(AGAUR,Catalunya) and by RFBR grant 06-01-00609 (Russia).

\ 

\noindent
{\bf Appendix} We now consider how the universe can reach the exit of the inflation. 
For simplicity, we consider the model in (\ref{Uf5}). In order to compare with usual scenario 
of the inflation, we work in the scalar-tensor form in the Einstein frame (\ref{XXX11}). 
In the epoch of the inflation, the curvature $R=A$ could be large and we find 
\be
\label{A1}
f(R) \sim - \frac{1}{f_1} + \frac{f_0}{f_1^2 R^{2n+1}}\ ,
\ee
and therefore
\be
\label{A2}
\sigma \sim \frac{(2n+1)f_0}{f_1^2 A^{2n+2}}\ ,\quad 
V(\sigma) \sim \frac{1}{f_1} - \frac{2(n+1)f_0}{f_1^2}\left(\frac{f_1^2 \sigma}{(2n+1)f_0}\right)^{\frac{2n+1}{2n+2}}\ .
\ee
Since $\sigma$ is now dimensionless, the condition for the slow roll could be given by $\left|V'/V\right|\ll 0$. 
Now we have
\be
\label{A3}
\frac{V'(\sigma)}{V(\sigma)} \sim - f_1 \left(\frac{f_1^2 \sigma}{(2n+1)f_0}\right)^{-\frac{1}{2n+2}}\ .
\ee
If we start with $\sigma \sim 1$, by using (\ref{Uf7}), we find
\be
\label{A4}
\frac{V'(\sigma)}{V(\sigma)} \sim - \left(\frac{R_0^{2n}}{\Lambda_i^{2n}}\right)^{\frac{1}{2n+1}}\ ,
\ee
which could be very small and the slow roll condition could be satisfied. 
The potential $V(\sigma)$ in (\ref{A2}) tells that if we start with $\sigma\sim 1$, the value of $\sigma$ increase very slowly 
and therefore $R$ becomes smaller. Then $\sigma$ becomes large enough, the inflation could stop.

\end{document}